\documentclass[12pt]{article}
\usepackage{amstex}
\usepackage{amssymb}

\advance\hoffset -0.5cm
\advance\voffset -1.0cm
\newlength{\height}
\divide \height by 3

\begin{document}


\def\spur#1{\mathord{\not\mathrel{#1}}} 
\baselineskip=\height 


\begin{titlepage}
\begin{center}

\makebox[\textwidth][r]{SNUTP-96-036}

\vskip 0.35in
{{\Large \bf Dynamical Mass in Strongly Coupled QED\footnote[1]{Lecture 
given at 14th Symposium on Theoretical Physics, ``Dynamical Symmetry
Breaking and Effective Field Theory", 21-27 July, 1995, Cheju, Korea}
}}
\end{center}
\begin{center}
\par \vskip .1in \noindent Deog Ki Hong\footnote[2]
{E-mail address: dkhong@@hyowon.cc.pusan.ac.kr} 
\end{center}
\begin{center}
Department of Physics, Pusan National University\\ 
Pusan 609-735, Korea\\

\par \vskip .1in 
\noindent
\end{center}

\begin{abstract}
We review dynamical mass generation in three and four dimensional
Ableian gauge theories. The basic approach analyzed here is to
solve Schwinger-Dyson equations in some approximations. In both
cases, the coupling has to be larger than a critical value  to have
dynamical symmetry breaking. In $QED_4$, chiral symmetry is
spontaneouly broken if $\alpha>\alpha_c (\simeq \pi/3)$ in ladder
approximation. The electron becomes massive, and is realized as 
Skyrmion in the low energy effective action. In
$QED_3$, electron gets dynmical mass in $1/N$ expansion if
$N<N_c(\simeq 32/\pi^2)$. With a Chern-Simons term, parity tends to
break maximally in $QED_3$. The quntum phase structure of $QED_3$
with or without a Chern-Simons term is peculiar.

\vskip 0.2in
\noindent


\end{abstract}

\end{titlepage}
\section{Introduction}
The idea of spontaneous symmetry breaking constitutes 
the essential part of the 
Standard Model which is proven to be a right theory to describe 
Nature upto several $100GeV$ energy scale. 
By the success of the idea, most of us believe 
that all different-looking forces in Nature can be explained 
by a single principle. 
One compelling way to break symmetry spontaneously is by dynamics.   
Namely, asymmetric ground state is dynamically favorable.  
We call this dynamical symmetry breaking (DSB) \cite{miranskybook}. 
It seems the only way of spontaneous symmetry breaking 
that is known to be realized in Nature so far. The examples are the 
BCS theory of superconductivity, superfluid $He^3$, chiral symmetry 
breaking in QCD, ${\it etc.}$   

It is obvious that we need strong coupling 
to achieve DSB.\footnote[1]{For BCS superconductivity, arbitrary weak 
attraction leads to instability of ground state due to the presence of 
Fermi surface.} For weak coupling, pertubation is good and 
vacuum in the pertubation theory respects the symmetry of Lagrangian. 
A naive explanation of DSB in strongly coupled system is following. 
Quantum fluctuation creates a virtual pair of fermion and  
antifermion. The pair  may form a bound state  
virtually.  But, if the binding energy of the bound state  
is larger than the energy needed to create such a pair at free state, 
the energy of the bound 
state will be lower than that of the pertubative vacuum. Then, it is
energetically preferable to pump  infinitely many bound states out of the 
perturbative vacuum, indicating the instability of the perturbative vacuum. 
The Bose condensate of the bound states defines a new vacuum,  
asymmetric if the condensate is not invariant under the symmetry of 
Lagrangian. The particle spectra will be quite different from the 
fields in the Lagrangian.  

For several reasons, though there is no direct proof, 
we believe that DSB occurs in QCD; the chiral symmetry of QCD is 
spontaneously broken by strong interaction.  
The reasons are, among others,  
(1) there are 8 light pseudoscalar mesons (pions and kaons), 
compared to $\Lambda_{\rm QCD}$, (2) lattice calculations\cite{seyong} 
show $\left<\bar\Psi\Psi\right>\ne0$, and 
(3) the anomaly argument of 
't Hooft-Coleman-Witten in $1/N$ \cite{coleman}.    
But, the full understanding of DSB in QCD is still lacking. 
Understanding how chiral symmetry is broken in QCD 
is important not only in hadron physics but also 
in the evolution of our Universe or in the proposed heavy ion collisoin 
experiment or in constructing models for dynamical 
breaking of electroweak 
symmetry. There are several models studied to understand DSB in a 
simplified setup but containing the general feature of QCD. They are 
Nambu-Jona-Lasinio model \cite{nambu}, strongly coupled quenched QED
\cite{qed4},  three dimensional large $N_f$ QED \cite{appel}, three
dimensional Thirring model \cite{thirring},  {\it etc.} 
All the models show that 
DSB occurs in a strong phase.  

In this lecture, I will concentrate on 
strongly coupled QED in four and three dimensions.  
As we will see later, in the strong phase of QED, 
a fermion bilinear operator gets large anomalous mass dimension and 
the wave function of the electron-positron bound state collapses 
due to the strong coupling in deep UV region, 
leading to spontaneous breaking of chiral symmetry. This phenomena   
of strongly coupled QED is used in several models for 
dynamical breaking of the electroweak symmetry as in the top-quark 
condensate model and in the walking technicolor model \cite{yamawaki}.

\section{Strongly coupled Quantum Electrodynamics}

Since in QED the vacuum polarization screens the electric charge at 
long distance, 
the electric charge becomes stronger and stronger in deep UV region, 
which leads to so-called ``Moscow zero" or ``Landau ghost"
\cite{landau}. It means that the renormalized electic charge has to
be zero  in order that QED may have well-defined limit as we remove
the UV cut-off of the theory; QED is a trivial theory.  
If QED is an effective theory below a scale, 
$\Lambda <\Lambda_{\rm QED}$,  
then we do not have to worry about the triviality 
of QED. 
Another way to avoid this triviality problem in QED is that 
QED has a nontrivial UV fixed point. The latter possibility was studied 
by several people, starting from the program of Baker, Johnson and Willey 
and Adler, known as finite QED \cite{baker}. 
Near a fixed point, if it exists, the running effect of the coupling 
will be unimportant, thus ladder approximation is applicable. 
Maskawa and Nakajima \cite{maskawa} studied the 
Schwinger-Dyson (SD) equation for the fermion propagator in the 
ladder approximation in QED.  They found chiral symmetry is 
spontaneously broken if the coupling is larger than 
a critical coupling, $\alpha>\alpha_c$. 
($\alpha_c={\pi\over3}$ in Landau gauge.) The explicit analysis 
for the SD-equation for the fermion propagator was done by Fukuda and 
Kugo \cite{fukuda}. 
Fomin, Gusynin and Miransky analyzed the Bethe-Salpeter (BS) 
equations for spinless bound state in quenched QED and showed that 
for a supercritical coupling the bound state wave function 
collapses, which then leads to chiral symmetry breaking \cite{fgm}.  

Bardeen, Leung and Love \cite{bardeen}argued that in 
the strong phase of QED a four-Fermi operator becomes 
relevant due to the large anolamous dimension of $\bar\psi\psi(x)$. 
Soon after, with adding the chirally invariant operator
\begin{equation}
{g^2\over \Lambda^2}\left[ \left( \bar\psi\psi\right)^2
+\left( \bar\psi\gamma_5\psi\right)^2\right],
\end{equation} 
it is found that there is a critical line in the coupling 
space of $(g^2, \alpha)$ for chiral symmetry breaking \cite{inoue}. 
Hong and Rajeev analyzed the solutions 
to the SD equations in quenched QED in the language of the operator 
product expansion and found that the 
solutions consistently describe spontaneous chiral symmetry 
breaking in the strong phase of QED \cite{hong1}.  
They also constructed an effective action for the strongly coupled 
QED with spontaneous chiral symmetry breaking, and argued that 
the bosonized QED admits a stable fermionic soliton, 
which can be identified as a massive electron,  and thus 
the strong QED does not exhibit confinement, even though 
chiral symmetry breaking does occur \cite{hong2}.   

\subsection{ The Schwinger-Dyson equation in the quenched QED}

We derive the gap equation for QED. One of the Schwinger-Dyson 
equations for QED provides the relation between the full fermion 
propagator, the photon propagator, and the three-point vertex, which is 
in euclidean notation
\begin{equation}
\spur{p} A(p^2)-B(p^2)=\spur{p}-m_0
 +e\int_k\gamma^{\mu}D_{\mu\nu}(p-k){-\spur{k} A-B\over A^2 k^2 +B^2} 
\Lambda^{\nu}
\end{equation}    

We approximate the photon propagator by free propagator in Landau 
gauge, and the vertex by the tree level value, $e\gamma^{\nu}$. 
The resulting equation then may be separated into two equations according to
the pieces with different spinor matrix structures.

\begin{eqnarray}
 B(p^2) &=&m_0+3e^2\int_k{1\over (p-k)^2}{B\over A^2k^2+B^2}  \\
 A(p^2)&=&1-{e^2\over p^2}\int_k{g_{\mu\nu}-(p-k)_{\mu}(p-k)_{\nu}/(p-k)^2
\over (p-k)^2}  {\rm tr}\spur{p}\gamma^{\mu}\spur{k}\gamma^{\nu} 
{A\over A^2 k^2+B^2}
\end{eqnarray}
Taking trace over the gamma martrices, the wave function renormalization 
constant becomes 
\begin{equation}
 A(p^2)=1+4{e^2\over p^2}\int_k \left( {p\cdot k\over (p-k)^2}
+2{p\cdot (p-k) k\cdot (p-k) \over (p-k)^4} \right){A\over A^2k^2+B^2}
\label{angular}
\end{equation}
Note that the angular integration of eq. (\ref{angular}) vanishes 
\begin{equation}
 \int_0^{\pi}\sin^2\theta d\theta 
    \left\{ {pk\cos\theta\over (p^2+k^2-2pk \cos\theta)}
    +2{(p^2-pk\cos\theta)(pk\cos\theta-k^2)\over 
    (p^2+k^2-2pk \cos\theta)^2}\right\}=0
\end{equation}
which would not be true if the gauge fixing parameter $\xi\ne \infty$. 
Therefore, for ladder approximation in Landau gauge, $A(p^2)=1$ and 
\begin{equation}
 B(p^2)=m_0+3e^2\int_k {1\over (p-k)^2}{B\over k^2 +B^2}
\label{gap}
\end{equation} 
Using the formular 
\begin{equation}
 \int_0^{\pi}\sin^2\theta d\theta {1\over (p^2+k^2-2pk\cos\theta)}
={\pi\over2pk}\left[ {k\over p}\theta(p-k)+
{p\over k}\theta(k-p)\right]
\end{equation}
eq. (\ref{gap}) becomes, after angular integration, 
\begin{equation}
 B(p^2)=m_0+{3\alpha\over 4\pi}\int dk^2
{k^2B\over k^2+B^2}\left[{\theta(p^2-k^2)\over p^2}+
{\theta(k^2-p^2)\over k^2} \right]
\label{gap1}
\end{equation}
One may convert the above integral equation into a differential 
equation. We differentiate eq.(\ref{gap1}) with repect to $p^2$ 
to obtain
\begin{equation}
 {d B\over d p^2}=-3{\alpha\over 4\pi(p^2)^2}\int_0^{p^2}dk^2
{k^2 B\over k^2+B}
\label{dif1}
\end{equation}
Multiplying $(p^2)^2$ and differentiating once again we obtain
\begin{equation}
 {d\over d(p^2)}\left( (p^2)^2{dB\over dp^2}\right)=
{-3\alpha\over 4\pi}{p^2B\over p^2+B^2}.
\end{equation}
Converting the integral equation into a differential equation, 
we get two boundary conditions:
\begin{eqnarray}
 \lim_{p^2\to0}\left( (p^2)^2{dB(p^2)\over dp^2}\right)&=&0  \\
\lim_{p^2\to\Lambda^2}\left[ p^2{dB(p^2)\over p^2}+B(p^2)\right]
&=&m_0, 
\end{eqnarray}
where $\Lambda$ is a UV cut-off.

Letting $p^2=x$ and $\omega=\sqrt{1-3\alpha/\pi}$, we find 
\begin{equation}
x{d^2B\over dx^2}+2{dB\over dx}+{(1-\omega^2)\over4}
    {B\over x+B^2} =0
\label{dif3}
\end{equation}
One may linearize the above equation as 
\begin{equation}
x{d^2B\over dx^2}+2{dB\over dx}+{(1-\omega^2)\over4}
        {B\over x+m^2} =0 
\end{equation}
where $m=B(0)$. This is a good approximation in both infrared 
$(p^2\ll m^2)$ and ultraviolet $(p^2\gg m^2)$ regions. The 
numerical analysis shows that it is good even at $p^2\sim m^2$.   

Putting $z=-x/m^2$ and $Y=B(x)/m$, we get 
\begin{equation}
z(1-z){d^2 Y\over dz^2}+(2-2z){dY\over dz}-{1\over4}(1-\omega^2)Y=0. 
\label{dif2}
\end{equation}
The general solution to the above equation (\ref{dif2}) is a 
linear combination of two independent hypergeometric functions
\begin{eqnarray}
Y(z)&=&C_1\:  {}_2F_1(a,b, 2; z)\\
&+&
C_2\left[ (-z)^{-a}{}_2F_1(a,-b,2b;z) 
+(-z)^{-b} {}_2F_1(b,-a,2b;z) \right], 
\end{eqnarray}
where $a=(1+\omega)/2$, $b=(1-\omega)/2$ and 
the hypergeometric function is in a series form 
\begin{equation}
{}_2F_1(a,b,c;z)={\Gamma(c)\over \Gamma(a)\Gamma(b)}
\sum_{n=0}^{\infty}{\Gamma(a+n)\Gamma(b+n)\over \Gamma(c+n)}{z^2\over n!}. 
\end{equation}
Since, as $p^2\to0$, $B\to m$, $Y(0)=1$. Therefore $C_1=1$ and 
$C_2=0$, namely 
\begin{equation}
B(p^2)=m\:{}_2F_1({1+\omega\over2}, {1-\omega\over2}, 2; -p^2/m^2)
\end{equation} 
The UV boundary condition yields a relation between 
$m$, $m_0$ and $\Lambda$;
\begin{eqnarray}
 m_0 &\simeq &m{\Gamma(\omega)\over \Gamma\left({\omega+1\over 2}\right) 
\Gamma\left({\omega+3\over2}\right)}
\left({\Lambda\over m}\right)^{\omega-1}; 
\quad{\rm at}\quad \alpha<{\pi\over3},   \\
m_0 &\simeq& m{4\over\pi}\left( \ln{\Lambda\over m}+2\ln2-1\right)  
\left( {\Lambda\over m}\right)^{-1};\quad{\rm at }\quad \alpha={\pi\over3} \\
m_0 &\simeq& {m^2\over \Lambda}\left( {2\coth(\pi\bar\omega)\over 
\pi\bar\omega}\right)^{1/2}\sin \left( \bar\omega\ln{\Lambda\over m}
+\theta(\bar\omega)\right);\quad{\rm at}\quad \alpha>{\pi\over 3},
\end{eqnarray}
where $\bar\omega=\sqrt{3\alpha/\pi-1}$ and 
$\theta(\bar\omega)=Arg \left( \Gamma(1+\omega)/
\Gamma^2((1+\omega)/2)\right)$. 
We see that for $\alpha\le\pi/3$ no dynamical mass is generated 
for a finite $\Lambda$ in the chiral limit ($m_0\to0$). On the 
other hand, for  $\alpha>\pi/3$, there is a dynamically generated mass 
if 
\begin{equation}
 \sin \left( \bar\omega\ln{\Lambda\over m}
+\theta(\bar\omega)\right)=0. 
\end{equation} 
Near the critical coupling, $\theta(\bar\omega)\simeq0$ and the 
dynamically generated mass is 
\begin{equation}
 m\simeq\Lambda \exp \left( -\pi n/\bar\omega\right), \quad n=1,2,\cdots.
\end{equation}
For $\Lambda\to\infty$, we want $m$ to remain finite. This is possible only 
if the coupling has a nontrivial dependence in $\Lambda$: for
$n=1$\footnote[2]{In fact, $n=1$ corresponds to a correct 
bound-state wavefunction.}  
\begin{equation}
 \alpha(\Lambda)=\alpha_c+{\alpha_c\pi^2\over 
\left[\ln{\Lambda\over\mu}\right]^2}, 
\label{run}
\end{equation}
where $\mu$ is some scale we introduced by hand as a dimensional 
transmutation (Note that if we remove the cut-off the theory has no scale.)  
The corresponding $\beta$ function is then 
\begin{equation}
 \beta(\alpha)\equiv\Lambda{\partial\over \partial\Lambda}\alpha=
-{2\over\pi\sqrt{\alpha_c}}(\alpha-\alpha_c)^{3/2}
\end{equation}
We see that the $\beta$ function is a nonanalytic function of the
coupling constant as indication that the origin of this $\beta$
function is nonperturbative.  As Miransky pointed out this
nonperturbative running coupling originates  from the wave function
collapse for the electron-positron  bound state when $\alpha>\alpha_c$
\cite{miransky}.

\subsection{Vacuum energy}

The Schwinger-Dyson equation is derived from the stationary condition 
for the effective action of a composite operator $G(x,y;K)$ 
for vanishing source $K$:
\begin{equation}
 {\delta \Gamma\over \delta G(x,y)}=0
\end{equation} 
Therefore, to find the true vacuum solution, one has to 
check the stability condition. 

Define an effective vacuum energy $V(G)$
\begin{equation}
 \Gamma(G)=-V(G)\int d^4x
\end{equation}
where we use the translational invariance of the two-point function. 
Then, the effective vacuum enegy is, for the quenched QED, 
\begin{eqnarray}
 V(G) &=& - {1\over 4\pi^2}\int p^2 dp^2  
\left[{1\over 2}\ln \left( 1+{B^2(p)\over p^2}\right)
+{p^2\over p^2+B^2(p)}-1 \right] \nonumber \\
 & - &{3e^2\over 128\pi^4}\int dp^2 dk^2
{p^2B(p) k^2B(k)\over \left[ p^2+B^2\right] \left[ k^2+B^2\right]}
\left\{ {\theta(p^2-k^2)\over p^2}+
{\theta(k^2-p^2)\over k^2}\right\}
\label{vac}
\end{eqnarray}
where we have normalized the vacuum energy such that $V(B=0)=0$. 
One can easily see that any nontrivial solution ($B(p)\ne0$) 
 has a lower energy 
than the trivial solution, $B(p)=0$. Therefore, chiral symmetry is 
spontaneously broken once chiral-symmetry breaking solution is found, 
which we have seen to exist for $\alpha>\pi/3$ in quenched QED.

\subsection{Nambu-Goldstone bosons and wave-function collapse}

For $N(\ge2)$ identical ``electrons", the Lagrangian has 
$SU(N)_L\times SU(N)_R\times U(1)_V$ symmetry. 
From the Ward-Takahashi identity for the axial current inserted in a 
two-point function, 
\begin{eqnarray}
 \partial_z^{\mu}\left<0\left|Tj_{\mu}^{\alpha}(z)\psi(x)\bar\psi(y)
\right|0\right>&=&{i\over2}\left<0\left|T\bar\psi(z)\{m_0,
\lambda^{\alpha}\}\psi(z) \psi(x)\bar\psi(y)\right|0\right> 
\nonumber \\
&+&i\delta^4(x-z)
\left<0\left|T \left(
i{\lambda\over2}\psi(x)\bar\psi(y)\right)\right|0\right> \\
&+&i\delta^4(x-z) \left<0\left|T
\left(\psi(x)\bar\psi(y)(-i{\lambda\over2})\right)\right|0\right>
\nonumber
\label{wt}
\end{eqnarray} 
where $\lambda^{\alpha}/2$'s are the generators for $SU(N)$,
one can derive in the massless limit a relation in momentum space 
\begin{equation}
 P^{\mu}\Gamma^{\alpha}_{5\mu}(p,k;P)=
-G^{-1}(p){\lambda^{\alpha}\over2}\gamma_5-
{\lambda^{\alpha}\over2}\gamma_5G^{-1}(k). 
\end{equation} 
where the vertex in momentum space is defined as 
\begin{equation}
 iG(p)\Gamma^{\alpha}_{5\mu}iG(k)\delta^4(p-k-P)
=\int dxdydz e^{ipx-iky-iPz}
\left<0\left|Tj_{5\mu}^{\alpha}(z)\psi(x)\bar\psi(y)\right|0\right>
\end{equation}  
Note near the pole 
\begin{eqnarray}
 \tilde\Gamma_{5\mu}(x,y)&=&\left<0\left|j_{5\mu}^{\alpha}(0)\right|P,
\beta\right>    {-i\over
P^2+i\epsilon}\left<P,\beta\left|\psi(x)\bar\psi(y)\right|0\right>
\nonumber \\ 
&=&iFP_{\mu}\delta^{\alpha\beta}{-i\over P^2+i\epsilon}
\chi^{P;\beta}(x,y)
\end{eqnarray}
where $F$ is the ``electropion" decay constant. 
When $p\to k$ (or $P\to0$)
\begin{equation}
 G(k)P^{\mu}\Gamma_{5\mu}^{\alpha}(k,k;P)G(k)=F\chi^{P;\alpha}(k)
\end{equation}
with 
\begin{equation}
 \chi(x,y)=\chi(x-y)=\int_k e^{ik\cdot (x-y)}\chi^{P;\beta}(k).
\end{equation}
One finds in the chiral limit 
\begin{equation}
 G^{-1}(q)\chi^{P;\alpha} G^{-1}(q)=
-\{G^{-1}(q),{\lambda^{\alpha}\over2}\gamma_5\}
\end{equation} 
Or, in the case at hand where $A(q)=1$, 
\begin{equation}
 \left(\spur{q}-B(q^2)\right)\chi^{P;\alpha}(q) 
\left( \spur{q}-B(q^2)\right)=\lambda^{\alpha}B(q^2)\gamma_5. 
\end{equation}
We may expand the amplitude for the bound state in the Dirac matrices
\begin{equation}
 \chi^{P;\alpha}(q)=\left[ \chi_1^{P}+\chi_2^{P}\spur{P}+\chi_3^{P}
\spur{q}+\chi_4^{P}\sigma^{\mu\nu}(P_{\mu}q_{\nu}
-P_{\nu}q_{\mu})\right]
{\lambda^{\alpha}\over2}\gamma_5.
\end{equation}
As $P\to0$, the pseudo-scalar channel survives to give 
\begin{equation}
 \chi_1^{P}(q)={2\over F}{B(q^2)\over q^2+B^2(q^2)}.
\end{equation}
We see that the electron self-energy is closely related to the 
bound-state wave function.

Without going into the WT identity, one can also see the relation 
directly from the 
Bethe-Salpeter (BS) equation;
\begin{equation}
 \left( {1\over2}\spur{P}+\spur{q}-m_a\right)_{nn_1}
    \chi^{P}(q)_{n_1m_1}
\left( {1\over2}\spur{P}-\spur{q}-m_a\right)_{m_1m}
=\int_q K_{mn,n_2m_2}(q-k)\chi^{P}(k)_{n_2m_2}.
\end{equation}
The BS kernel is given as 
\begin{equation}
 K_{mn,n_2m_2}(q-k)=e^2 \left( \gamma^{\mu}\right)_{nn_2}
    D_{\mu\nu}(q-k) \left( \gamma^{\nu}\right)_{m_2m}
+e^2 \left( \gamma^{\mu}\right)_{nm} 
\left( \gamma^{\nu}\right)_{m_2n_2}D_{\mu\nu}(P),
\end{equation}
where we have included the dynamically gernerated fermion mass, 
$m_a$. 

For the pseudo-scalar channel with $P\to0$ one gets 
\begin{equation}
 (m^2+q^2)\chi^{P}_{ab;1}(q)=
{12\pi\alpha}\int_k{1\over(q-k)^2}\chi^{P}_{ab;1}(k). 
\label{bs1}
\end{equation}
where we take $m_a=m_b=m$.  
(More precisely, we should take $B(q^2)$ for $m_a$ and 
$m_b$, but taking $m$ for fermion self energy is a good approximation 
in the region we are interested in.)
Note that the second term in the kernel, corresponding to  
annihilation process, does not contribute to the BS equation in 
ladder approximation. If we let 
\begin{equation}
 (m^2+q^2)\chi^{P}_{ab;1}(q)={\rm const.}\times B(q^2),
\end{equation}
we see that $B(q^2)$ satisfies the linearized gap equation
\begin{equation}
 B(q^2)={12\pi\alpha}\int_k{1\over (q-k)^2}{B(k^2)\over k^2+m^2}
\end{equation}
This equation has been solved exactly. But, here we solve it again in 
the position space to find the physical meaning of the oscillating 
solutions for a supercritical coupling. 
The BS equation eq.(\ref{bs1}) is after Fourier-transforming to 
the position space
\begin{equation}
 (m^2-\partial^2)\chi^{P}(x)={3\alpha\over\pi}{1\over r^2} \chi(x)
\label{bs2}
\end{equation}
where $r=\sqrt{x_ix_i}$. 
Eq.(\ref{bs2}) can be thought of as a Schr\"odinger equation 
for a bound-state in 
four-dimensional euclidean space with an attractive potential 
\begin{equation}
 \left( -\partial^2- {3\alpha\over\pi}{1\over r^2}\right)\chi^P(x)
=-m^2 \chi^P(x)
\end{equation} 

Since large $m\left(\simeq B(0)\right)$ corresponds to the solution of
lower vacuum  energy to the gap equation, we look for the ground state
of the Schr\"odinger equation, eq. (\ref{bs2}). Because of the scale
invariance, we see that if  $\chi^P(x)$ is a solution to the
Schr\"odinger equation with eigen value $-m^2$, so is $\chi^P(\lambda
x)$ with eigen value $-m^2/\lambda^2$. Therefore the spectrum of the
Schr\"odinger equation  is continuous. If there is a bound state, the
ground state energy has to be $-\infty$. In quantum mechanics, we know 
that most of time particle exists in the region where the
energy of the particle is larger than the potential energy. Thus, the 
ground state wave function collapses toward the origin. To make sense
out of this divergence, we regularize the potential as follows:
\begin{equation}
V(r)=
\begin{cases}
-{\alpha/\alpha_c\over r^2},& \text{if $r\ge a$},\\
-{\alpha/\alpha_c\over a^2},& \text{otherwise}, 
\end{cases}
\end{equation}
where $a$ is a short-distance cut-off and $\alpha_c={\pi\over3}$. At
short distance, $r\ll 1/m$,  one may neglect the $m^2$ term. Then, 
the $O(4)$-invariant solution satisfies
\begin{equation}
{\chi^P}^{\prime\prime}(r)+{3\over r}{\chi^P}^{\prime}(r)
-V(r)\chi^P(r)=0.
\end{equation} 
When $r\ge a$, 
\begin{equation}
\chi^P(r)=A r^{s_1}+Br^{s_2}
\end{equation}
with $s_1=s_2^*=-1+i\sqrt{\alpha/\alpha_c-1}$. Similarly, when
$0\le r\le a$,

\begin{equation}
\chi^P(r)={C\over r}J_1(kr),
\end{equation}
where $C$ is a constant, $k=\sqrt{\alpha/\alpha_c}\>a^{-1}$, and 
$J_1$ is the Bessel function of the first kind. The boundary condition 
that both of $\chi^P(r)$ and its derivative be continuous 
at $r=a$ leads to 
\begin{equation}
{B\over A}=e^{2i\theta}a^{s_1-s_2}
\end{equation}
with
\begin{equation}
\theta=Arg\left(\sqrt{\alpha/\alpha_c-1}+i\sqrt{\alpha/\alpha_c}
{J_1^{\prime}(\sqrt{\alpha/\alpha_c})\over 
J_1(\sqrt{\alpha/\alpha_c})}\right). 
\end{equation}
Then, for $a\le r\ll 1/m$, 
\begin{equation}
\chi^{P}(r)={\tilde A\over r}\cos\left[\sqrt{\alpha/\alpha_c-1} 
\ln \left({r\over a}\right)+\theta\right],
\end{equation}
where $\tilde A$ is a constant. Now, if we let $a$ go to zero, the 
wave function oscillates rapidly and the number of zeros increases 
exponentially, which implies the energy of the bound state is 
$-\infty$ as we argued earlier. To regularize this divergence, we have
to let the coupling constant run as we change the cut-off such that 
\begin{equation}
\lim_{a\to0}\left[
\sqrt{\alpha(a)/\alpha_c-1}\ln({r\over a})+\theta(a)\right]=0. 
\end{equation}
Then we get a running coupling constant same as we obtained earlier
by the analysis of the Schwinger-Dyson equation for the electron
propagator;
\begin{equation}
\alpha(a)=\alpha_c+{\alpha_c\pi^2\over \left[
\ln(a\mu)\right]^2},
\end{equation}
where $\mu$ is the renormalization point.

\subsection{Operator Product Expansion}

Let us go back to the differential gap equation, eq. (\ref{dif3}). 
For large momentum, $p\gg B(p^2)$, the equation is linearized as 
\begin{equation}
 p^2 {d^2B\over dp^2}+3p {dB\over dp}+rB=0
\label{linear1}
\end{equation}
where $r=3\alpha/\pi$. 
When $r<1$, the solutions are for $p\gg B(p)$  
\begin{equation}
 B(p)\simeq m_R \left( {\mu\over p}\right)^{\epsilon}
+{\kappa\over p^2} \left( {\mu\over p}\right)^{-\epsilon}
\label{sol1}
\end{equation}
The operator product expansion of the propagator in momentum space is  
\begin{equation}
\lim_{p^2\to\infty} \left<\psi\bar\psi(p)\right>={A(p/\mu)\over \spur{p}} 
+{C(p/\mu)m_R(\mu)\left<1\right>\over p^2} 
+{D(p/\mu)\left<\bar\psi\psi\right>\over p^4}+\cdots
\label{ope}
\end{equation}
where $A,C$ and $D$ are the coefficient functions. 
If we plug the solutions to OPE, 
\begin{eqnarray}
 \lim_{p^2\to\infty} \left<\psi\bar\psi(p)\right>&=&{1\over \spur{p}}+
{B(p)\over p^2}+\cdots\\
&=&{1\over \spur{p}}+{m_R\over p^2}\left( {\mu\over p}\right)^{\epsilon}
+{\kappa\over p^4}\left( {\mu\over p}\right)^{-\epsilon}+\cdots
\end{eqnarray}
We finds that $m_R$ is the renormalized mass and 
$\kappa=\left<\bar\psi\psi\right>$. As analyzed by Cohen and Georgi, 
one can show 
that when $r<1$ there is no solution $\kappa\ne0$ in the chiral 
limit $m_R\to0$. For $r<1$, $m_R$ is never zero except for the trivial 
solution, $B(p)=0$. 

When $r>1$, the solutions to the gap equation at large momentum are 
\begin{equation}
 B(p)={\kappa\over p}\sin \left[ \sqrt{r-1}\ln(p/\mu)+\varphi\right]
\end{equation} 
where $\kappa$ and $\varphi$ are two parameters that characterize the 
solution. Now, if we take the non-perturbative running of coupling 
eq. (\ref{run}) proposed by Miransky, 
we get 
\begin{equation}
 B(p)={\bar\kappa\over p}
\end{equation}
where $\bar\kappa=\kappa\sin(\pi+\varphi)$. We see that,  
for $r>1$ with the non-perturbative running coupling, two operators 
$m_R$ and $\bar\psi\psi$ coalesce; 
also, the anomalous dimension of $\bar\psi\psi(x)$ becomes 1. 
But, as shown earlier, for 
$r>1$ we can take $m_0\to0$ with keeping $m=B(0)$ finite and thus 
$\bar\kappa\ne0$. Therefore, one can conclude that the solution found 
for the quenched QED with non-perturbative running of coupling 
consistently describes spontaneous breaking of chiral symmetry 
in the language of operator product expansion.  

\subsection{Electron as a soliton in bosonized QED}
In this section we describe the low energy effective action of the 
strong phase of QED and argue that massive electrons appear 
as skyrmionic solutions in the effective action \cite{hong2}. 
Given that chiral symmetry is spontaneously broken in the strong 
phase of QED, one may try to construct an low energy effective action 
for the strong phase. At low energy, the right degrees of freedom are 
massless Nambu-Goldstone bossons of chiral-symmetry breaking 
and massless photon. The electrons are massive and decoupled. 

We define 
\begin{equation}
U^i_j(x)=\lim_{x\rightarrow y}{\left|x-y\right|\over \kappa}
\bar q^i(x)\left( 1+i\gamma_5\right)q_j(y),
\end{equation} 
which transforms as   
\begin{equation}
U(x)\longrightarrow g_LU(x)g_R^{\dagger},
\end{equation} 
where $g_R$ and $g_L$ are $N\times N$ unitary matrices. 
Here, $N(>1)$ is the number of flavors. 

In a vacuum $U$ has a constant (nonzero) expectation value, which 
we can choose to be $\left<U_j^i\right>=\delta_j^i$. Then, 
the Nambu-Goldstone bosons are described by 
\begin{equation}
U(x)=g_L^{\dagger}(x)g_R(x),
\end{equation} 
which satisfies $U^{\dagger}U=1$. 
Since the current corresponding to the field ${\rm det}U$ is the 
$U_A(1)$ current which is anormalous due to the Adler-Bell-Jackiw anomaly, 
we must impose ${\rm det}U=1$ and we have $N^2-1$ massless 
Nambu-Goldstone bosons. 

To the lowest order (in derivative expansion) in the effective action
is the sum of their kinetic terms:
\begin{equation}
S_{\rm eff}=\int dx \left( {1\over4}F_{\mu\nu}F^{\mu\nu}+
{1\over 2}F^2{\rm tr}\partial_{\mu}U^{\dagger}\partial_{\mu}U\right)
+\cdots.
\label{eff}
\end{equation}  
Here, $F$ is the electropion decay constant. 
The effective action (\ref{eff}) has too much symmetry. It is invariant 
under two separate discrete symmetries. As Witten\cite{witten} 
has argued in the context 
of QCD, one has to add the Witten-Wess-Zumino term\cite{current} 
in the effective action 
in order for the effetive theory has same symmetry as QED:  
\begin{equation}
S_{\rm WWZ}=i{5n\over 240\pi^2}\int dx\;d\tau\epsilon^{\mu\nu\rho\sigma}
{\rm tr}{\tilde U}^{-1}{\partial\tilde U\over \partial\tau}
{\tilde U}^{-1}\partial_{\mu}{\tilde U}{\tilde U}^{-1}
\partial_{\nu}{\tilde U}{\tilde U}^{-1}\partial_{\rho}{\tilde U}
{\tilde U}^{-1}\partial_{\sigma}{\tilde U}.
\label{wwz}
\end{equation}
The coefficient $n$ of the Witten-Wess-Zumino term has to be integer 
in order for the effective action to be consistent at quantum level. 
Futhermore, by calculating the three-pont function, whose residue at the 
$p^2=0$ pole is determined by the Adler-Bell-Jackiw anomaly,  
\begin{equation}
\left<J_{\mu5}(p)J_{\nu5}(q)J_{\rho5}(r)\right>
\end{equation}
both in the effective action  and in QED, we get $n=1$. 

Though the Nambu-Goldstone bosons are electrically neutral, 
it must have coupling with photons, since they are made of 
charged particles. The coupling term with lowest number of derivatives is 
\begin{equation}
\beta \int dx A_{\mu}J^{\mu},
\end{equation} 
where $J^{\mu}$ is the topological current 
\begin{equation}
J^{\mu}={1\over 24\pi^2}\epsilon^{\mu\nu\rho\sigma}{\rm tr}U^{-1}
\partial_{\nu}UU^{-1}\partial_{\rho}UU^{-1}\partial_{\sigma}U. 
\end{equation}
The charge of the topologiacl current corresponds to the winding 
number of the third homotopy group of $SU(N)$. 
The coefficient $\beta$ can be determined by comparing the pole at $p^2=0$ 
in the four-point function 
\begin{equation}
\left< {\rm tr} \left[ J_{\mu5}(p)J_{\nu5}(q)J_{\rho5}(r)A_{\sigma}(s)
\right]\right> 
\end{equation}
in QED and in the effective action. We get $\beta=e$, the 
electric charge. 

The final form of our effective action is then
\begin{equation}
S_{\rm eff}=\int dx \left[ {1\over4}F_{\mu\nu}F^{\mu\nu}+
{1\over2}F^2 {\rm tr}\partial_{\mu}U^{\dagger}\partial^{\mu}U 
+e A_{\mu}J^{\mu}\right] + S_{\rm WWZ}.
\label{eff2}
\end{equation}

We look for a solitonic excitation in the effective action.
We take the following Ansatz for the static and finite-energy solution 
of the winding number 1:
\begin{equation}
U_c(x)=e^{i\vec\tau\cdot \hat x \theta(r)}, \quad A_0=\omega(r), 
\quad A_i=0, 
\end{equation} 
where $\vec\tau$ is the generator of $SU(2)$ subgroup of $SU(N)$. 

The energy of this static configuration is 
\begin{equation}
E(\omega, \theta)=\int 4\pi r^2\;dr \left[ -{1\over2} 
{\omega^{\prime}}^2+F^2 \left( {\theta^{\prime}}^2+2{\sin^2\theta\over 
r^2}\right)+{e\over 2\pi^2}{\omega\over r^2}\sin^2{\theta}\theta^{\prime}
\right].
\end{equation}
Using Gauss' theorem, $\omega^{\prime}=Q(r)/4\pi r^2$, one can eliminate 
$\omega$ in terms of $\theta$. Then, one can get a lower bound on 
the energy, which is anlogous to the Bogolmolny bound for monopoles.
We find 
\begin{equation}
E>{\pi\over \sqrt{2}}eF.
\end{equation}
The upper bound can be obtained using the variational method. 
We get 
\begin{equation}
E<3.3eF.
\end{equation}
Therefore, we find there is a static configuration of finite energy. 
Now, we try to find its quantum number. To this end, we have to 
quantize the zero modes of the soliton. 

If $U(x)$ is a solution, so is $AUA^{-1}$ for any $A\in SU(N)$. 
$A_1$ and $A_2$ are equivalent if $A_1=A_2h$ and $h$ is in the commutant 
$U(N-2)$ of $SU(2)$ in $SU(N)$. Therefore, we see that $A$ belong to 
the coset space $M=SU(N)/U(N-2)$. Right multiplication 
of $A$ by $h\in SU(2)$ corresponds to spatial rotations 
and the left multiplication by $g\in SU(N)$ to flavor transformations. 

The effective Lagrangian for the zero modes $A$ can be obtained by 
substituting $U(\vec x, t)=A(t)U_c(\vec x)A(t)^{-1}$ into 
(\ref{eff2}):
\begin{equation}
L[A]=-E+{1\over2}I_{\alpha\beta}{\rm tr}\lambda_{\alpha}A^{-1}
\dot A{\rm tr} 
\lambda_{\beta}A^{-1}\dot A -i{1\over2}{\rm tr}YA^{-1}\dot A, 
\end{equation}
where $E$ is the energy of the static soliton, $I_{\alpha\beta}$ is an
invariant tensor on $M$ and $Y$ is a hypercharge,
\begin{equation}
Y=
\begin{pmatrix}
    1 & 0 & 0\\
    0 & 1 & 0\\
    0 & 0 & 0_{(N-2)\times(N-2)}
\end{pmatrix}
-{1\over N}
\begin{pmatrix}
    1 & 0 & 0\\
    0 & 1 & 0\\
    0 & 0 & 1_{(N-2)\times(N-2)}
\end{pmatrix}.
\end{equation}
Under the transformation $A(t)\rightarrow A(t)h(t)$ with 
$h\in U(N-2)$
\begin{equation}
L\rightarrow L-i{1\over2} {\rm tr}Yh^{-1}\dot h.
\end{equation}
It follows that the wavefunction of the soliton must satisfy 
\begin{equation}
\begin{split}
\Psi(Ah)&=\Psi(A)\quad {\rm if}\quad h\in SU(N-2)\\
\Psi(AH)&=e^{i(1-2/N)\theta}\Psi(A)\quad{\rm if}\quad h=e^{iY\theta}.
\end{split}
\end{equation}
The simplest solutions are 
\begin{equation}
\Psi(A)=D_{ab}(A),
\end{equation}
where $D_{ab}(A)$, $a=1, \cdots, N$ and $b=1,2$ is the matrix representing 
$A$ in the fundamental representation of $SU(N)$. These wave functions form a
doublet under the right action of $SU(2)$. Therefore, the ground 
state of the soliton is a spin-half 
particle transforming under the fundamental representation of the flavor
group. Futher more, the particle must be a fermion, since after $2\pi$ 
rotation the wavefunction changes sign. We conclude that the soliton in the 
effective Lagrangian of QED is the electron. Its physical mass is 
\begin{equation}
{\pi\over \sqrt{2}}eF<m<3.3eF.
\end{equation}

\vfill
\pagebreak 

\section{Large $N$ QED in three dimensions}
Three dimensional gauge theories have been studied intensively 
recently, since they can describe the high temperature 
limit of 4-dimensional field theories, as well as their relevance 
to planar condensed matter system.  

Quantum electrodynamics in 2+1 dimensions ($QED_3$) is 
super-renormalizable and has severe IR divergence, which is 
softened in $1/N$ expansion, where $N$ is the number of flavors. In the 
IR region, the dimensionful coupling, $\alpha$, drops out  and $1/N$ 
behaves like a dimensionless coupling in $1/N$ expansion of $QED_3$. 
The dynamical symmetry breaking in $QED_3$ has been studied 
in the framework of the Schwinger-Dyson equation by Appelquist {\it 
et. al.}\cite{appel} They showed for even $N$ that, 
when $N<N_c\simeq 32/\pi^2$, dynamical fermion 
mass is generated in the chiral limit and the flavor symmetry 
$U(2N)$ breaks down  to $U(N)\times U(N)$:
the dynamical mass is obtained as 
\begin{equation}
 m\sim \alpha\exp \left[ {-2\pi\over \sqrt{N_c/N-1}}\right],
\end{equation} 
where $\alpha\equiv e^2/N$. 
This result was supported by a lattice calculation\cite{kogut}. 
But, there has been a criticism on this work because using 
the bare vertex together with $A(p)=1$ may lead to a result inconsistent  
with $1/N$ expansion\cite{pennington}. 
To overcome this shortcoming, there has been an attempt to use a
(nonlocal) gauge where $A(p)=1$ exactly and use an Ansatz for the
vertex function to satisfy  the WT identity\cite{kondo}. They found 
a finite critical number of flavors, $N_c=128/3\pi^2$. 
 
On ther other hand, 
Pisarski \cite{pisarski} argued, using an effective field theory
approach with $\epsilon(=4-d)$ expansion, 
that for any $N$ the dynamical fermion mass is generated, 
namely, the critical number of flavor is $\infty$.  

In this section, I will discuss the dynamical mass generation in
$QED_3$ and the property of the phase transition near $N_c$, since it 
has some resemblance to four-dimensional 
quenched QED. At the end 
I will review the effect of Chern-Simons term on the generation 
of dynamical mass and the phase transition.

\subsection{Schwinger-Dyson equation in $QED_3$}

The Lagrangin of three dimensional quantum
electrodynamics\footnote[3]{ Here we discuss a parity-invariant
theory}   is  given as 
\begin{equation}
 {\cal L}=\sum_{i=1}^N 
\bar\psi_i(i\spur{\partial}-e\spur{A})\psi -{1\over 4} 
F_{\mu\nu}^2
\end{equation}
where $\psi$ is a four-component spinor. 
The Lagrangian has $U(2N)$ symmetry generated by 
\begin{eqnarray}
J_{a}^{\mu}&=\bar\psi\gamma^{\mu}T_a\psi, \quad
J_{a5}^{\mu}&=\bar\psi\gamma^{\mu}T_a\gamma^5\psi
   \\ 
J_{a3}^{\mu}&=\bar\psi\gamma^{\mu}\gamma^3T_a\psi, \quad
J_{a35}^{\mu}&=\bar\psi\gamma^{\mu}T_a\gamma_3\gamma^5\psi
\end{eqnarray}
To have a well-defined theory for large $N$ we keep $\alpha=Ne^2/8$ 
fixed as $N\to\infty$. 
The photon propagator in $1/N$ expansion is obtained by 
summing up all the 
bubble diagrams, which is in Landau gauge 
\begin{equation}
 D_{\mu\nu}(p)={g_{\mu\nu}-p_{\mu}p_{\nu}/p^2\over 
    p^2 \left[ 1+\Pi(p)\right]}
\label{photon}
\end{equation}
where 
\begin{equation}
 \Pi(p)={\alpha\over 8 p}. 
\end{equation}

The leading order Schwinger-Dyson equation is in euclidean notation 
\begin{eqnarray}
 A(p) & = & 1 + \left( {1\over N}\right)\\
 B(p) &=  & {8\alpha\over N}\int_k
{\gamma^{\mu}D_{\mu\nu}(p-k)B(k)\gamma^{\nu}\over k^2 +B^2(k)}
\label{gap_31}
\end{eqnarray}
Angular integration in eq. (\ref{gap_31}) gives 
\begin{equation}
 B(p)={4\alpha\over \pi^2 Np}\int dk {kB(k)\over k^2 + B^2(k)}
\ln \left[ {k+p+\alpha\over |k-p|+\alpha}\right]
\end{equation}
using the formular 
\begin{equation}
 \int_0^{\pi}\sin\theta d\theta{1\over (p^2+k^2-2pk\cos\theta)\ 
\left(1+{\alpha\over \sqrt{p^2+k^2-2pk\cos\theta} } \right)}=
{1\over pk}\ln \left[ {p+k+\alpha\over |p-k|+\alpha}\right]
\label{gap_32}
\end{equation}
Since $QED_3$ is super-renormalizable, the integral eq. (\ref{gap_32}) 
is rapidly damped for $p>\alpha$. For $p<\alpha$, 
it takes the approximate form 
\begin{equation}
 B(p)={4\over\pi^2 Np}\int_0^{\alpha} dk {k B(k)\over k^2+ B^2}
(k+p-|k-p|)  
\label{gap_33}
\end{equation}
where $\alpha$ is used as a UV cut-off and the leading term in 
the logarithm is kept:
\begin{equation}
\ln \left[ {p+k+\alpha\over |p-k|+\alpha}\right]=
\left[ {p+k\over \alpha}- 
{|p-k|\over\alpha}\right] 
\end{equation}
Eq. (\ref{gap_33}) can be replaced by the differntial equation 
\begin{equation}
 {d\over dp} \left[ p^2{dB(p)\over dp}\right]=-\left[{8\over \pi^2N} \right]
{p^2 B(p)\over p^2 +B^2(p)}
\label{gap_34}
\end{equation}
together with 
the boundary conditions
\begin{equation}
 0\le B(0)<\infty
\end{equation}
and 
\begin{equation}
 \lim_{p\to \alpha} 
\left[ p{dB(p)\over dp}+B(p)\right]=0
\label{bc_32}
\end{equation}
The boundary condition eq.(\ref{bc_32}) 
insures the absence of a bare mass.  
In a regime $p\gg B(p)$, as in the case of quenched $QED_4$, 
the eq. (\ref{gap_34}) is linearized 
by replacing $B(p)$ with $B(0)$ in the 
denomenator to give a power solution
\begin{equation}
 B(p)\sim p^a
\end{equation} 
where 
\begin{equation}
 a=-{1\over 2}\pm{1\over2}\sqrt{1-{32\over \pi^2N}}
\end{equation}
For $N>32/\pi^2$, the power solution does not satisfy the 
boundary condition eq. (\ref{bc_32}). But, 
for $N<32/\pi^2$, it does and the fermion self energy is oscillatory,  
\begin{equation}
 B(p)={\kappa\over \sqrt{p}}\sin \left[ {1\over2} 
\sqrt{32/ \pi^2N-1} \ln \left( {p\over B(0)}\right) +\varphi\right], 
\end{equation}
which satisfies  eq. (\ref{bc_32}) if 
\begin{equation}
{1\over2} 
\sqrt{32/ \pi^2N-1} \ln 
\left( {\alpha\over B(0)}\right) +\varphi 
=n\pi
\end{equation}
Thus we find  
\begin{equation}
 B(0)\simeq\alpha\exp 
\left[ {-2(n\pi-\varphi)\over \sqrt{32/\pi^2N-1}}\right]
\end{equation}
We see that for $N<N_c= 32/\pi^2$
the fermions get dynamical mass which vanishes smoothly 
as $N\to N_c$. 
It looks like the phase transition is of second-order.

\subsection{Quantum Phase Transition in $QED_3$} 
Pisarski studied an effective field theory which has same symmetry as
$QED_3$ with $N$ massless four-component spinors:
\begin{equation}
 {\cal L_{\rm eff}}={1\over2} tr \left( \partial_{\mu}\phi\right)^2 
+{8\pi^2\mu^{\epsilon}\over 4!}
\left\{ g_1 \left[ tr(\phi^2)\right]^2 +g_2 tr(\phi^4)
 \right\}
\label{effective}
\end{equation}
where all renormalizable terms are kept but  
all cubic couplings and and quadratic couplings are absent      
in the effective Lagrangian,
assuming the multiciritical point.
$\phi$ is a traceless $N\times N$ Hermitian matrix and 
tranforms under $SU(N)$ as 
\begin{equation}
 \phi\to U^{\dagger}\phi U.
\end{equation} 
Note that the trace degree of freedom, which is parity-odd, 
is massive and decoupled  at the 
multiciritiacl point since 
it can not have v.e.v due to Vafa-Witten theorem \cite{vafa}.  

He showed that, for $N>\sqrt{5}/2$, there are no IR stable fixed points 
by analyzing the $\beta$-functions for $g_1$ and $g_2$ obtained by 
$\epsilon$-expansion. He then concluded that mass must be generated  
and thus fermions are massive for all $N$. Therefore, if the effective 
theory described by eq.(\ref{effective}) is in a universality class 
with $QED_3$, dynamical fermion mass is generated for all $N$ 
in $QED_3$. 

To examine the basic assumptions of Pisarski argument, Appelquist, Terning 
and Wijewardhana \cite{atw} studied the fermion-antifermion bound
state in the symmetric phase of $QED_3$. They solved the
Schwinger-Dyson equations  for fermion-antifermion scattering
amplitude in $1/N$ expansion.  The Schwinger-Dyson equation
is  
\begin{eqnarray}
 T_{\lambda\rho\sigma\tau}(p,q) & =  &  {16\alpha\over N}
\left( \gamma^{\mu}\right)_{\sigma\lambda}D_{\mu\nu}(p) 
\left( \gamma^{\nu}\right)_{\rho\tau}\\
 &+& \int_k T_{\lambda\rho\sigma^{\prime}\tau^{\prime}}(k,q) 
{1\over {1\over2}\spur{q}+\spur{k}} 
\left( \gamma^{\mu}\right)_{\sigma\sigma^{\prime}}D_{\mu\nu}(p-k)
\left( \gamma^{\nu}\right)_{\tau^{\prime}\tau}
{1\over {1\over2}\spur{q}+\spur{k}} 
\end{eqnarray}
The scattering amplitude can be written as 
\begin{equation}
 T_{\lambda\rho\sigma\tau}(p,q)=\delta_{\lambda\rho}\delta_{\sigma\tau} 
T(p,q)+\cdots
\end{equation} 
where the ellipsis indicates pseudoscalar, vector, axial vector, and
tensor components. The scalar channel factorizes in the SD equationsi
and gives 
\begin{equation}
 T(p,q)={16\alpha\over 3Np(p+\alpha)}
+{16\alpha\over 3\pi^2Np}\int {dk\over k} T(k,q) \ln 
\left[ k+p+\alpha\over |k-p|+\alpha\right]. 
\end{equation} 
Since we are interested in the massless bound-state at the critiical point 
we try to see if there exists a pole as $q\to0$. Since 
$QED_3$ damps rapidly as 
$p\gg \alpha$ and $q$ acts as an infrared cutoff, we use the approxiamtion:
\begin{equation}
 T(p,q)={16\over 3Np} +{16\over 3\pi^2 Np}
\int_q^{\alpha} {dk T(k,q)\over k} \left( k+p-|k-p|\right). 
\label{pole} 
\end{equation} 
For $p>q$, eq. (\ref{pole}) can be converted to a differential equation:
\begin{equation}
 p{d^2\over dp^2}\left( pT\right)=-{32\over 3\pi^2 N}T
\label{pole1}
\end{equation}
The solutions of eq. (\ref{pole1}) are 
\begin{equation}
 T(p,q)={A(q)\over\alpha} 
\left( {p\over \alpha}\right)^{-{1\over2}+{1\over2}\eta}
+{B(q)\over\alpha}
\left( {p\over \alpha}\right)^{-{1\over2}-{1\over2}\eta}
\end{equation}
where 
\begin{equation}
 \eta=\sqrt{1-N_c/N}
\end{equation}
and $N_c=128/3\pi^2$. The coefficent $A$ and $B$ are determined as 
\begin{eqnarray}
 A &=  &{-\left( {1\over2}-{1\over2}\eta\right)^2 \pi^2 
\left( {q\over \alpha}\right)^{-{1\over2}+{1\over2}\eta}
\over 
2 \left( {1\over2}+{1\over2}\eta\right)
\left( 1- \left( {1-\eta\over 1+\eta}\right)^2 
\left( {q\over \alpha}\right)^{\eta}\right)},\\
B & = &
{\left( {1\over2}-{1\over2}\eta\right) \pi^2 
\left( {q\over \alpha}\right)^{-{1\over2}+{1\over2}\eta}
\over 
2 \left( 1- \left( {1-\eta\over 1+\eta}\right)^2 
\left( {q\over \alpha}\right)^{\eta}\right)}. 
\end{eqnarray}
The amplitude has a pole at $q_0$ in the complex $q$ plane with 
\begin{equation}
 |q_0|=\alpha \left( {1+\eta\over 1-\eta}\right)^{2\over\eta}
\end{equation}
In the limit $\eta\to0$, 
\begin{equation}
 |q_0|\to \alpha e^4
\end{equation}
We see that there is no pole at the critical point. Thus the assumption of 
the existence of muticritical point in Pisarki's analysis 
is not applicable in the case of large $N$ $QED_3$. The phase transition 
near $N_c$ seems quite peculiar: in the broken phase the dynamical 
fermion mass, being an order parameter for the phase transition, 
vanishes smoothly as we approach the critical point, while in the 
symmetric phase the mass of the bound state does not vanish near 
the critical point. 

\subsection{$QED_3$ with a Chern-Simons term}
In 2+1 dimensional gauge theories  
the gauge bosons can have parity-violating mass by  
Chern-Simons term for the gauge fields. 
Since the Chern-Simons term breaks parity, it induces 
not only parity-violating mass to fermion but also fractional spin. 
Here I will briefly discuss the effect of the Chern-Simons term in 
dynamical generation of parity-even mass and also in the phase
transition of $QED_3$.  As studied by Hong and Park \cite{hp}, if
Chern-Simons term is added in $QED_3$,  it tends to break parity
maximally. Namely, the dynamical parity-even  fermion mass gets
reduced in the magnitude and  the critical coupling ($1/N_c$) 
gets larger. Later, Kondo and Maris \cite{koma} analyzed this model
in a nonlocal gauge for which $A(p)=1$ and found similar results. 
Analyzing the model 
for small coefficient of Chern-Simons term, 
They showed the dynamical parity-even fermion mass 
does not vanish as $N_c\to N$, indicating that the phase transition is 
first order in contrast to the phase transition in pure $QED_3$. 

One can study the fermion-antifermion scattering amplitude 
in the symmetric phase of $QED_3$ in the presence of Chern-Simons term, 
following the analysis of Appelquist {\it et. al} for pure $QED_3$
\cite{hong}. Using the coefficient of the Chern-Simons term as an
expansion parameter, it is found that the mass of the scalar channel
bound state vanishes as we approach the critical point, which shows
that in the presence of Chern-Simons term $QED_3$ exhibits a
peculiar phase transition as in the case of pure $QED_3$.

\section{Conclusion}
As we have seen in the analysis of Abelian gauge theory in three and
four dimensions, chiral symmetry is dynamically broken when the
coupling is larger than a critical coupling. 

In four dimensions, the  critical coupling  is $\pi/3$ in Landau
gauge and it turns out to be the UV fixed point. 
The $\beta$-function is due to the wave-function collapse of
the electron-positron bound state. Quenched $QED_4$ exhibits two-phase
structure. In the strong phase, chiral symmetry is spontaneously
broken and the low energy spectrum contains the massless
Nambu-Goldstone bosons, massless photon, and massive electrons as 
we have seen in the low energy effective field theory of $QED_4$. The
massive electron is nothing but a Skyrmion in the bosonized quantum
electrodynamics.

In three dimensions, $1/N$ expansion is a sensible way to study
nonperturbative phenomena in $QED_3$, since it is super
renormalizable and has severe IR divergence in ordinary coupling
perturbation. If one uses $1/N$ expansion, $1/N$ becomes an effective
coupling of electron and positron in IR region and 
dynamical electron mass
generates when $1/N$ is larger than a critical value (or $N<N_c$). 
The quantum phase structure of $QED_3$ seems peculiar; The order
parameter apporaches zero at the critical point in the asymmetric
phase but the correlation length remains finite at the critical point
in the symmetric phase.

With a Chern-Simons term, the parity of $QED_3$ tends to break
maximally. Namely, the magnitude of parity-even electron mass 
and the critical number of flavors get reduced. In the asymmetric
phase the order parameter remains finite at the critical point while
the correlation length gets infinite in the symmetric phase.

\vskip .1in
\noindent

{\bf Acknowledgments}
\vskip .1in

This work was supported in part by the Korea Science and Engineering 
Foundation through SRC program of SNU-CTP, 
and also by Basic Science Research
Program, Ministry of Education, 1995 (BSRI-95-2413).

\pagebreak

\end{document}